\documentclass[final,3p,times,twocolumn]{elsarticle}
 \biboptions{comma,sort&compress}
\usepackage{here}
 \usepackage{graphicx}
  \usepackage{epsfig}
\def\nin{\noindent}
\def\beq{\begin{equation}}
\def\eeq{\end{equation}}
\def\bea{\begin{eqnarray}}
\def\eea{\end{eqnarray}}

\journal{Nuc. Phys. (Proc. Suppl.)}

\begin{document}

\begin{frontmatter}

\title{Novel QCD Phenomena at the LHC: The Ridge, Digluon-Initiated Subprocesses, Direct Reactions, Non-Universal Antishadowing, and Forward Higgs Production}

 \author[label1]{Stanley J Brodsky\corref{cor1}}
  \address[label1]{SLAC National Accelerator Laboratory, Stanford University\\
   Stanford, California}
\cortext[cor1]{Speaker}
\ead{sjbth@slac.stanford.edu}


\begin{abstract}
\noindent
I discuss a number of novel tests of QCD at the LHC, measurements which can illuminate fundamental features of hadron physics. I also review the ``Principle of Maximum Conformality" (PMC) which systematically sets the renormalization scale order-by-order in pQCD, eliminating an unnecessary theoretical uncertainty. The PMC allows LHC experiments to test QCD much more precisely, and the sensitivity of LHC measurements to physics beyond the Standard Model is increased.

\end{abstract}


\end{frontmatter}



\section{Introduction}
 
The LHC has the unique capability to  test novel and unexpected aspects of hadron and nuclear physics predicted  by QCD and the Standard Model at very high energies.  In this short review I will focus on several examples: (a) the production of `ridge'  correlations in high multiplicity proton-proton collisions  and its connections to the dynamics controlling color confinement;  (b) digluon-initiated subprocesses -- a novel multiparton hadroproduction mechanism which can explain the observed nuclear suppression of  quarkonium production at forward rapidity;  
(c) flavor-dependent  antishadowing of the nuclear structure functions; and (d)  high $x_F$ Higgs hadroproduction, which is  predicted from the existence of intrinsic heavy-quark Fock states in the proton's light-front wavefunction.

Hadronic processes measured at the LHC involving high momentum transfer probe the structure and compositeness of quarks and gluons at unprecedented short distances. They  can test for the existence of 
new types of particles carrying color  such as new flavors of quarks and supersymmetric partners.  Precision tests of QCD  such as the measurements of the top pair production cross section and the $t \bar t$ asymmetry are complimentary to the search for new fields and particles.  As I will discuss,  the "Principle of Maximum Conformality" (PMC)~\cite{Brodsky:2013vpa,Brodsky:2011ig}.  
 provides a systematic and unambiguous way to set the renormalization scale and effective number of quark flavors in pQCD predictions at each order of perturbation theory.  An unnecessary  uncertainty from QCD theory is eliminated.  The PMC thus allows the LHC to test QCD much more precisely, and the sensitivity of LHC measurements to physics beyond the Standard Model is greatly increased. 

The versatility of the LHC in testing fundamental aspects of hadron physics can be extended  by 
adding  fixed-target capability, as in the AFTER (A Fixed-Target ExperRiment). proposal~\cite{Lansberg:2013wpx}, or by adding a high energy electron ring as in the LHeC proposal~\cite{AbelleiraFernandez:2012cc}.  In addition, by adding detectors which can measure particles  at high forward and backward rapidity,  an important new range of   phenomena can be explored; especially for testing predictions for  Higgs production at $x_F> 0.8$~\cite{Brodsky:2007yz,Brodsky:2006wb}.   

In the following sections I will discuss a number of LHC physics topics which test novel aspects of QCD.  Further details can be found in the references.

\section{The Origin of the  ``Ridge"  in Proton-Proton Collisions and the Connection to the Dynamics Underlying Confinement }

The CMS collaboration~\cite{Khachatryan:2010gv} at the LHC has reported a surprising phenomenon in very high-multiplicity high energy proton-proton collisions: a positive correlation between particles produced over a large rapidity interval along the same azimuthal angle as a trigger particle.  It had been previously believed that such ridge-like correlations would only occur in nucleus-nucleus collisions due to their elliptical overlap.    Bjorken, Goldhaber and I ~\cite{Bjorken:2013boa} have suggested that the  ``ridge"-like correlation  in $p p$ collisions reflects the rare events generated by the collision of aligned flux tubes which connect the valence quarks in the wave functions of the colliding protons. 
The ``spray" of particles resulting from the approximate line source produced in such inelastic collisions then gives rise to events with a strong correlation between particles produced over a large range of both positive and negative rapidity.   The highest multiplicity events will appears when the flux tubes have maximal overlap.  In the case when the $Y$-junctions between the three quarks of each proton overlap, one could also generate a $v_3$ pattern.

The physics of colliding flux tunes can also be studied in high multiplicity, high energy electron-ion collisions at the  proposed LHeC.  In this case the flux tube between the quark and antiquark of the virtual photon is oriented in azimuthal angle with the electron's scattering plane, and its characteristics -- such as its size in impact space  -- can be controlled by the photon's virtuality, as well as  the $q \bar q$ flavor.

The flux tube can be identified physically as the near-planar manifestation of the exchange of soft gluons which produce the fundamental color-confining interaction.  In fact, the entire Regge spectrum of light-quark baryons in $n$ and $L$ is well reproduced by the eigenvalues of a frame-independent light-front Shr\"odinger and Dirac equations with a confining quark-diquark potential dictated by the soft-wall AdS/QCD approach and light-front holography~\cite{Brodsky:2014yha,Brodsky:2013ar,Brodsky:2006uqa}.  Remarkably the same confining light-front potential arises from principle of
de Alfaro, Fubini, and Furlan ~\cite{de Alfaro:1976je} which allows a mass scale an a confining harmonic oscillator potential to appear in equations of motion without affecting the conformal invariance of the action.

The use of Dirac's light-front time $\tau=t+z/c$ provides a Lorentz frame-independent formulation of hadronic collisions, where the quark and gluon composition of hadrons or their formation at the amplitude level  is determined by their light-front wavefunctions -- the eigensolutions of the QCD light-front Hamiltonian~\cite{Bakker:2013cea}.  In this boost-invariant formalism, the light-front wavefunctions of the colliding protons (or ions) are frame-independent;  there is no  Lorentz contraction nor ``colliding pancakes"! 
This description of hadron dynamics also predicts the shape of the light-front wavefunctions $\psi_{n,H}(x_i, k_{\perp i})$  of the hadrons which underly hadronic form factors, structure functions, transverse momentum distributions, fragmentation functions, etc.  In addition, diffractive vector meson electroproduction is predicted successfully without any new parameter~\cite{Forshaw:2012im}.

\section{Higgs production at High $x_F$  and the Intrinsic Heavy-Quark Distributions of the Proton }

A first-principle prediction of QCD  is the existence of proton Fock states such as $|uud Q\bar Q>$. The light-wave function for such five-quark states have two origins:  (1) the usual DGLAP  ``extrinsic" contribution  arising from gluon splitting producing heavy quarks primarily at small  momentum fractions $x=k^+/P^+$, and (2) the ``intrinsic" contribution 
where the $Q \bar Q$ pair is multi-connected  to the proton's valence quarks~\cite{Brodsky:1980pb, Brodsky:1984nx}.  The intrinsic contributions arise for example from the cut of  the quark loop contribution to the  $g g \to g g$ amplitude in the wavefunction describing proton's self energy.  The  amplitude for the intrinsic contribution is maximum at minimum off-shellness where the  constituents tend have the same rapidity; i.e. $x_i \propto \sqrt{m^2_i + k^2_{\perp i}}$.   Thus the intrinsic heavy quarks in the Fock state carry most of the hadron's momentum. Since the effective four-gluon operator is twist-six, the operator product expansion predicts the probability 
for the intrinsic five-quark state is proportional to $1/m^2_Q$ in non-Abelian QCD~\cite{Brodsky:1984nx, Franz:2000ee}. The intrinsic contribution also leads to asymmetries in the $s$ vs. $\bar s$ momentum and spin distributions~\cite{Brodsky:1996hc}.

\begin{figure}
 \begin{center}
\includegraphics[height=6 cm,width=9 cm]{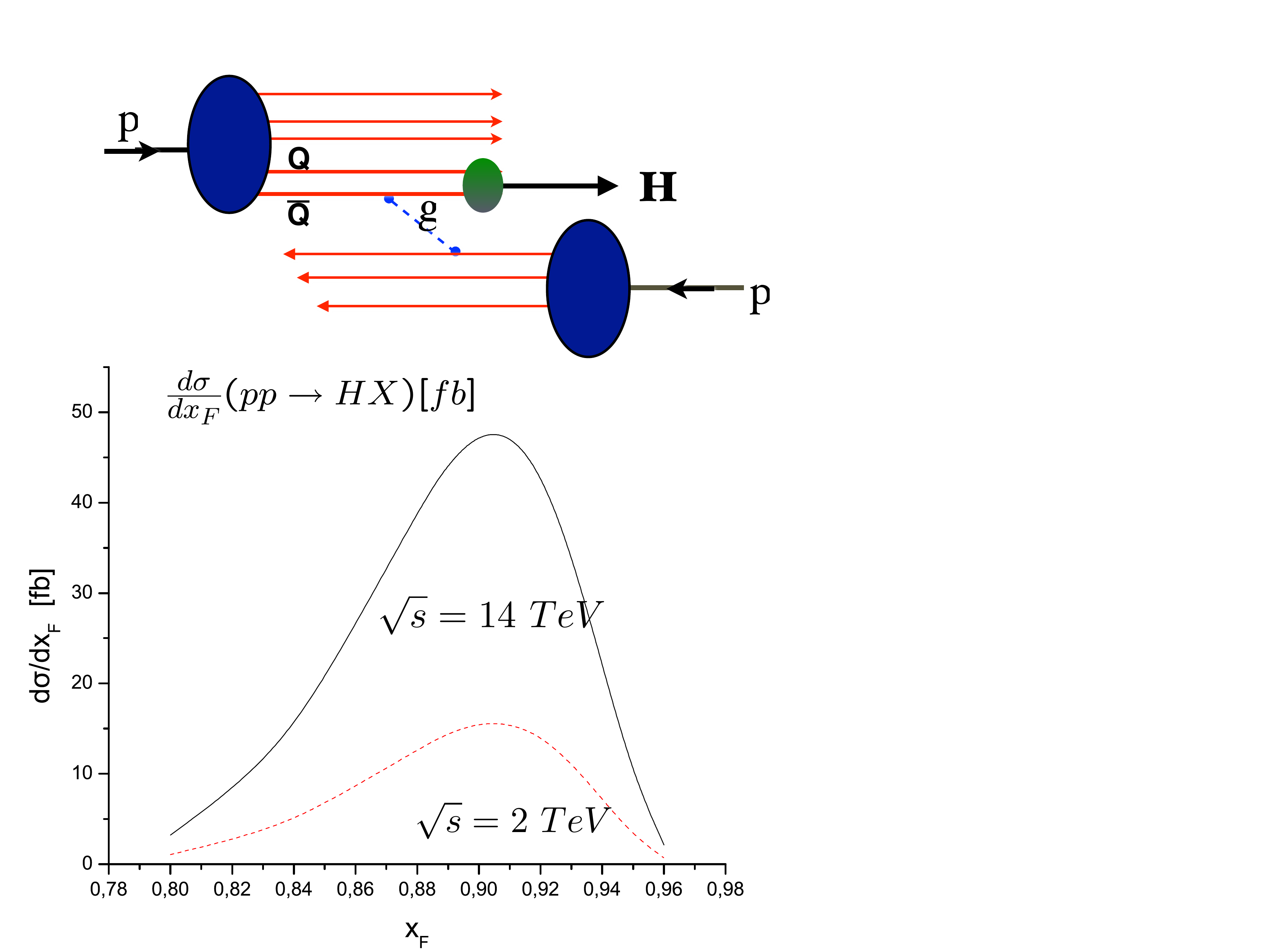}
\end{center}
\caption{Intrinsic Heavy Quark Mechanism and cross section for Higgs Production at LHC and Tevatron energies.} 
\label{Higgs}  
\end{figure}

The conventional pQCD mechanisms for Higgs production at the LHC such as gluon fusion $g g \to H$ lead to Higgs boson production in the central rapidity region.    
However the Higgs can also be produced at very high $x_F$ by the process $[Q\bar Q ] + g \to H$~\cite{Brodsky:2007yz}, where both heavy quarks from the  proton's five quark Fock state $|uud Q\bar Q>$ couple directly to the Higgs. See. fig. \ref{Higgs}. Since the Higgs couples to each quark proportional to its mass,  one has roughly equal contributions from intrinsic $s \bar s, c\bar c, b \bar b$ and even $t \bar t$ Fock states.  The intrinsic heavy-quark distribution of the proton at high $x$ leads to Higgs production 
with as much as $80\% $  of the beam momentum. The same mechanism produces the $J/\psi$ at high $x_F$ as observed in fixed-target experiments such as NA3.   The decay of the high-$x_F$ decay Higgs to muons could be observed using very forward detectors at the LHC. The predicted  cross section ${d\sigma/d x_F}( p p \to H X) $ for Higgs production at high $x_F \sim 0.8$ computed by Kopeliovitch, Schmidt, Goldhaber, and myself ~\cite{Brodsky:2007yz} is of order of 50 fb.  We have also computed with Soffer~\cite{Brodsky:2006wb} the corresponding double-diffractive rate $p p \to H  p p  X$.   Testing these  predictions would open up a new domain of Higgs physics at the LHC.

Signals for the Intrinsic heavy quark distribution at high $x$  have been seen by many experiments, including the measurement of $c(x,Q^2)$ at high $x$by the EMC collaboration and forward production at high $x_F$ of the $\Lambda_c$ at the ISR and SELEX at FermiLab,  $ \Lambda_b$, at the ISR,  double-charm baryons by SELEX, and the high $x_F$ production of the $J\psi$ and double $J/\psi $  by NA3  at CERN at FermiLab.  A review and references are given in  ref. ~\cite{Brodsky:2014bla}.
Since the $[Q\bar Q]$ is a color-octet $8_C$, the production of the $J/\psi$ at high $x_F$ from intrinsic charm in $p A$ collisions occurs at the nuclear front surface, explaining the observed $A^{2/3}$ nuclear dependence.  Intrinsic charm in the reaction $g c \to \gamma c$ has the potential to explain the anomalously large rate seen by D0 $p \bar p \to c \gamma X$ at high $p^\gamma_T$. Intrinsic charm is included in CTEQ 6.6M.  A signal for Intrinsic strangeness $s(x) + \bar s(x)$  in the region $0.1 < x< 0.4  $ from HERMES has been discussed by Chang and Peng~\cite{Chang:2011du}.  The duality of the $|uud s\bar s>$ Fock state with hadronic channels such as $|K |lambda>$ implies that the $s(x) $ and 
$\bar s(x)$ distributions will be different in shape~\cite{Brodsky:1996hc}.

\section{Digluon-Initiated Quarkonium Production}
Since the $J/\psi$-nucleon cross section is only a few millibarns,   the usual expectation is that the $J/\psi$ production cross sections in nuclei will be  approximately linear in the number of nucleons $A$.   However, the production cross section $ p A \to J/\psi X$ measured by LHCb~\cite{Aaij:2013zxa} and ALICE~\cite{Aamodt:2010jd} at forward rapidity $y \sim 4$  shows unexpectedly strong nuclear  suppression.    This strong suppression cannot be accounted for by shadowing of the nuclear gluon distribution.  

Arleo and Peigne~\cite{Arleo:2012hn,Arleo:2012rs}  have suggested that the strong nuclear suppression of $J/\psi$ production in $pA$ collisions can be explained as a manifestation of the ``color-octet" model:  the $c \bar c$ propagates through the nucleus as a color-octet, and its energy loss will be proportional to its energy if  the induced gluon radiation is coherent on the entire nucleus.  The color-octet $c \bar c$ is assumed to convert to the color-singlet  $J/\psi$ after exciting the nucleus.  


There is an alternative QCD mechanism for producing the $J/\psi$ in proton-nucleus collisions at  forward rapidity and small transverse momentum; digluon-initiated  subprocesses: $[gg] g \to J/\psi$.  Here the $[gg]$ is a color-octet digluon originating from the colliding proton; e.g. from its $|uud gg>$ Fock state.  
See fig. \ref{DiGluonAdep}.
It should be noted that since $g(x,Q^2)$ falls rapidly, two gluons in the digluon,  each with $x \sim 0.02$, could have a higher probability than a single gluon with $x \sim 0.04$.  The  digluon mechanism  will be suppressed at  high $p_T$,  but can dominate at low $p_T$ and forward rapidity.  The propagating color-octet digluon has a large interaction cross section, and it thus interacts primarily at the nucleus front surface, giving a production cross section  $\sigma(p A \to J/\psi X) \propto A^{2/3}.$   The produced $J/\psi$ then propagates essentially freely though the nuclear interior. The $\Upsilon$ produced by di-gluons will have a similar $A$ dependence.  However, this is not the case for the $\psi^\prime$; it can be further suppressed as it propagates through the nuclear environment  because of its larger radius.  
This digluon subprocess is the color-octet analog of the color-singlet  two-gluon exchange mechanism~\cite{Brodsky:2002ue} underlying diffractive processes  such as $\ell p \to \ell p X$.    
The digluon multiparton subprocess is  also analogous to the higher-twist subprocess $[q\bar] q q \to \gamma^* q $ which  dominates the 
$\pi N \to  \ell \bar \ell X$  Drell-Yan reaction at  high $x_F$ and is known to accounts  for the observed dramatic change from transverse to longitudinal virtual photon polarization~\cite{Berger:1979du}.  Similarly,  multiparton ``direct" subprocesses can account~\cite{Arleo:2009ch} for the observed anomalous power-law fall-off of high $p_T$ inclusive hadron production cross sections ${E d \sigma/^3p}(p p \to h X)$ at fixed $x_T = 2{p_T/ \sqrt s} $ and fixed  $\theta_{CM}$ as discussed in the next section.

The $ pA \to J/\psi X$ cross sections measured in fixed-target experiments at CERN and FermiLab at high $x_F$ also show a very strong nuclear suppression at high $x_F$.  The ratio of the nuclear and proton target cross sections has the form $A^{\alpha(x_F)}$, where $x_F$ is Feynman fractional longitudinal momentum of the $J/\psi$. At small $x_F$, $\alpha(x_F)$  is slightly smaller than one, but at $x_F \sim 1$, it decreases to $\alpha=2/3$. These results  are again surprising since (1) the dependence $A^{\alpha}= A^{2/3}$ is characteristic of a strongly interacting hadron, not a small-size quarkonium state; and (2) the functional dependence   
$A^{\alpha(x_F)}$ contradicts  pQCD factorization~\cite{Hoyer:1990us}.
\begin{figure}
 \begin{center}
\includegraphics[height=5cm,width=8cm]{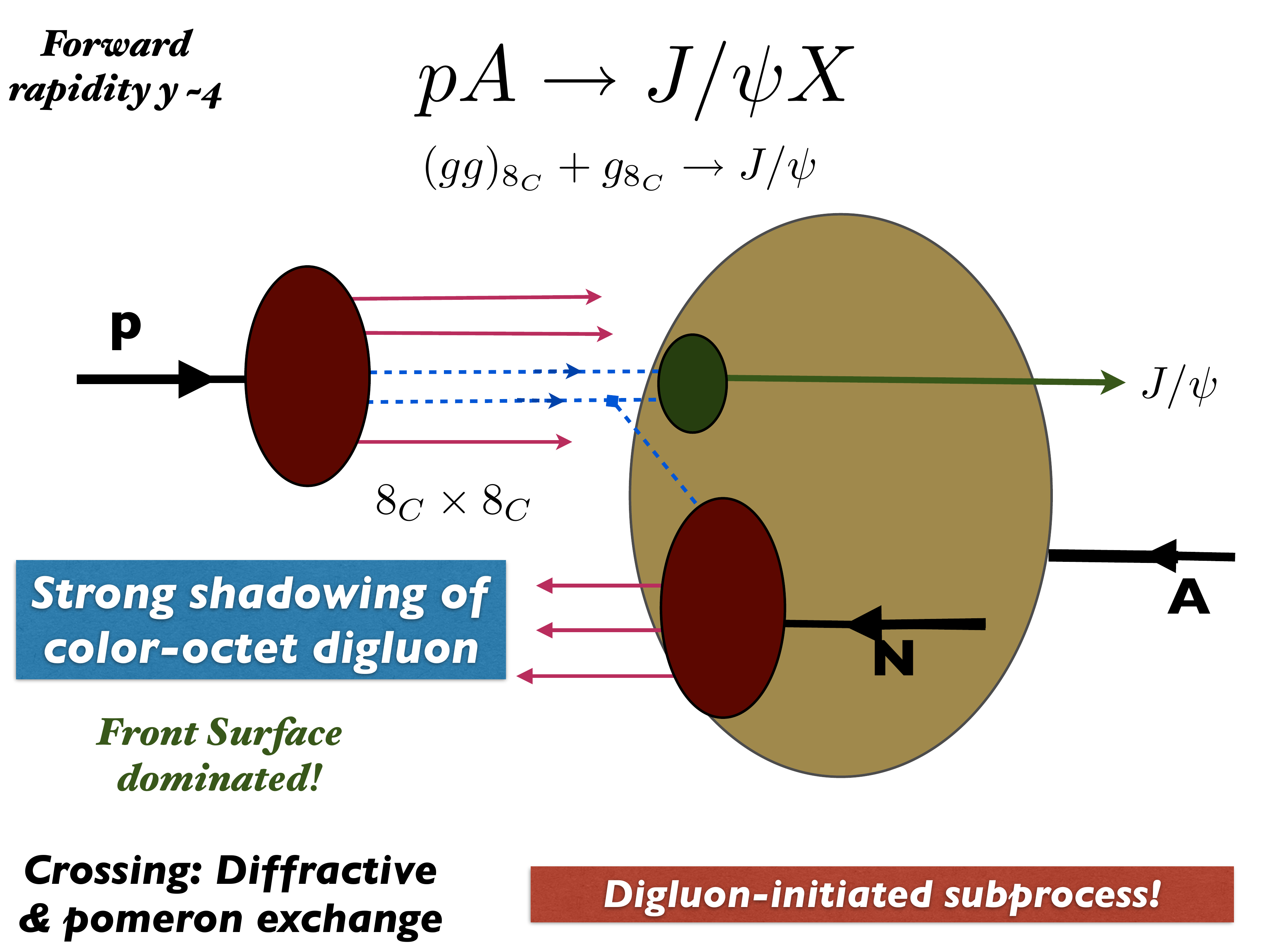}
\end{center}
\caption{Digluon-induced quarkonium production.  Since the color-octet digluon is strongly shadowed,  the production process occurs dominantly on the front surface of the nuclear target.} 
\label{Digluon}  
\end{figure} 
The observed nuclear suppression, in combination with the anomalously nearly flat cross section at high $x_F$ can be explained by the properties of 
the intrinsic charm Fock state~\cite{Brodsky:2007yz,Brodsky:2014bla}.
QCD predicts that the color-configuration of the heavy quark pair $Q \bar Q$ in the intrinsic five-quark Fock state  is primarily a color-octet.
The intrinsic heavy quark Fock state of the proton: $|(uud)_{8_C} (c \bar c)_{8_C} \rangle$ thus interacts primarily with the $A^{2/3} $ nucleons at the front surface because of the large color-dipole moment of the color-octet $c \bar c$.    The color-singlet quarkonium  state is thus produced at the front surface, and it then propagates through the nucleus with high $x_F$.

\section{The Unexpected Role of Direct Processes in High $p_T$ Hadron Reactions}

The factorization picture derived from the parton model and then from  pQCD has played a guiding role in virtually all aspects of hadron physics phenomenology.  In the case of inclusive reactions such as $  p p \to \pi X $,  the pQCD ansatz predicts that the cross section at leading order in pion's transverse momentum  $p_T$ can be computed by convoluting the perturbatively calculable hard subprocess  cross section with the process-independent structure functions and quark fragmentation functions.  
It is thus usually assumed that hadrons produced at high transverse momentum  in inclusive high energy hadronic collisions  arise dominantly  from jet fragmentation.
A  fundamental test of leading-twist QCD predictions in high transverse momentum hadronic reactions is the measurement of the power-law
fall-off of the inclusive cross section~\cite{Sivers:1975dg}
${E d \sigma/d^3p}(A B \to C X) ={ F(\theta_{cm}, x_T)/ p_T^{n_{\rm eff}} } $ at fixed $x_T = 2 p_T/\sqrt s$
and fixed $\theta_{CM}$. In the case of the scale-invariant  parton model  $n_{\rm eff} = 4.$    However in QCD,  $n_{\rm eff} \sim 4 + \delta$ where  $\delta \simeq 1.5 $  is the typical correction to the conformal prediction arising
from the QCD running coupling and the DGLAP evolution of the input parton distribution and fragmentation
 functions.~\cite{Arleo:2009ch,Arleo:2010yg}
The usual expectation then is that leading-twist subprocesses (i.e., the leading power-law contributions) will dominate high $p_T$ hadron production at RHIC and  at Tevatron energies.  Measurements of isolated photon production $ p p \to \gamma_{\rm direct} X,$ as well as jet production, do agree well with the leading-twist scaling prediction $n_{\rm eff}  \simeq 4.5$.~\cite{Arleo:2009ch}
However,   measurements  of  $n_{\rm eff} $ for hadron production  show much faster fall-off in $p_T$ at fixed $x_T$  and $\theta_{CM}$ and are inconsistent with the leading twist predictions.
Striking deviations from the leading-twist predictions were also observed at lower energy at the ISR and  Fermilab 
fixed-target experiments.~ \cite{Sivers:1975dg,Cronin:1973fd,Antreasyan:1978cw} This deviation points to a significant contribution from `direct'  higher-twist processes where the hadron is created directly in the hard subprocess, rather than from quark or gluon jet fragmentation.

Normally many more pions than protons are produced at high transverse momentum in hadron-hadron collisions. This is also true for the peripheral collisions of heavy ions. However, when the nuclei collide with maximal overlap (central collisions) the situation is reversed -- more protons than pions emerge.  This observation at RHIC~\cite{Adler:2003kg}  contradicts the usual expectation that protons should be more strongly absorbed than pions in the nuclear medium.

In fact, a significant fraction of high $p^H_\perp$ isolated hadrons can emerge
directly from hard higher-twist subprocess~\cite{Arleo:2009ch,Arleo:2010yg} even at the LHC.  
The direct production of hadrons can also explain~\cite{Brodsky:2008qp} the remarkable ``baryon anomaly" observed at RHIC:  the ratio of baryons to mesons at high $p^H_\perp$,  as well as the power-law fall-off $1/ p_\perp^n$ at fixed $x_\perp = 2 p_\perp/\sqrt s, $ both  increase with centrality,~\cite{Adler:2003kg} opposite to the usual expectation that protons should suffer more energy loss in the nuclear medium than mesons.
The high values $n_{\rm eff}$ with $x_T$ seen in the data  indicate the presence of an array of higher-twist processes, including subprocesses where the hadron enters directly, rather than through jet fragmentation.~\cite{Blankenbecler:1975ct}   Although they are suppressed by powers of $1/p_T$, the direct  higher twist processes can dominate because they are energy efficient -- no same-side energy or momentum is lost from the undetected fragments. Thus the incident colliding partons are evaluated at the minimum possible values of light-front momentum fractions $x_1$ and $x_2$, where the parton distribution functions are numerically large. Since these processes  create color-transparent baryons with minimal absorption, this mechanism can explain
the RHIC baryon anomaly.~\cite{Brodsky:2008qp}

\section{\bf Flavor-Dependent Antishadowing}

It has been conventional to assume that the nuclear modifications to the structure functions measured in deep inelastic charged lepton-nucleus and neutrino-nucleus interactions are identical. The antishadowing of the nuclear structure functions is particularly interesting. Empirically, one finds $F_{2A}(x,Q^2)/ (A/2) F_{d}(x,Q^2)
> 1 $ in the domain $0.1 < x < 0.2$;  { i.e.}, the measured nuclear structure function (referenced to the deuteron) is larger than the
scattering on a set of $A$ independent nucleons.  

One can show~\cite{Stodolsky:1994ka} using Gribov-Glauber theory  that the Bjorken-scaling diffractive deep inelastic scattering events lead to the shadowing of nuclear structure functions at small $x_{\rm Bjorken}.$  This is due to the destructive interference of two-step and one step amplitudes in the nucleus.   Since diffraction involves rescattering, one sees that shadowing and diffractive processes  are not intrinsic properties of hadron and nuclear wavefunctions and structure functions, but are properties of the complete dynamics of the scattering reaction.~\cite{Brodsky:2008xe} In fact,  Gribov-Glauber theory also predicts that the antishadowing of nuclear structure functions is not  universal, but depends on the quantum numbers of each struck quark and antiquark.~\cite{Brodsky:2004qa}  This could  explain the recent observation of Schienbein {\it et al.}~\cite{Schienbein:2008ay} who find that the nuclear structure functions  in the range $0.1 < x < 0.2$  measured by NuTeV in deep inelastic  neutrino  charged-current reactions differ significantly from the distributions measured in deep inelastic electron and muon scattering.    Note that there are leading-twist diffractive contributions $\gamma^* N_1 \to (q \bar q) N_1$  arising from Reggeon exchanges in the
$t$-channel. For example, isospin--non-singlet $C=+$ Reggeons contribute to the difference of proton and neutron
structure functions, giving the characteristic Kuti-Weiskopf $F_{2p} - F_{2n} \sim x^{1-\alpha_R(0)} \sim x^{0.5}$ behavior at small $x$. The
$x$ dependence of the structure functions reflects the Regge behavior $\nu^{\alpha_R(0)} $ of the virtual Compton amplitude at fixed $Q^2$ and
$t=0.$ The phase of the diffractive amplitude is determined by analyticity and crossing to be proportional to $-1+ i$ for $\alpha_R=0.5,$ which
together with the phase from the Glauber cut, leads to {\it constructive} interference of the diffractive and nondiffractive multi-step nuclear
amplitudes.  The coherence length only needs to be long enough to ensure coherence between the one-step and two-step Glauber processes, not the entire nuclear length.  The nuclear structure function is predicted~ \cite{Brodsky:1989qz} to be enhanced precisely in the domain $0.1 < x <0.2$ where
antishadowing is empirically observed.  Since quarks of different flavors couple to different Reggeons, this leads to the remarkable prediction that
nuclear antishadowing is not universal;~\cite{Brodsky:2004qa}  it depends on the quantum numbers of the struck quark. This picture implies substantially different
antishadowing for charged and neutral current reactions, thus affecting the extraction of the weak-mixing angle $\theta_W$.

\section{\bf Leading-Twist Lensing Corrections and the Breakdown of Perturbative QCD Factorization}

The effects of the final-state interactions of the scattered quark in deep inelastic scattering  have been traditionally assumed to either give  an inconsequential phase factor or power-law suppressed corrections.  However, this is only true for sufficiently inclusive cross sections.  For example, consider semi-inclusive deep inelastic lepton scattering (SIDIS) on a polarized target $\ell p_\updownarrow \to H \ell' X.$  In this case the final-state gluonic interactions of the scattered quark lead to a  $T$-odd non-zero spin correlation of the plane of the lepton-quark scattering plane with the polarization of the target proton~\cite{Brodsky:2002cx} which is not power-law suppressed with increasing virtuality of the photon $Q^2$; i.e. it Bjorken-scales.    This  leading-twist  ``Sivers effect"~\cite{Sivers:1989cc} is nonuniversal in the sense that pQCD predicts an opposite-sign correlation in Drell-Yan reactions relative to single-inclusive deep inelastic scattering.~\cite{Collins:2002kn,Brodsky:2002rv}
This important but  yet untested prediction occurs because the Sivers effect in the Drell-Yan reaction is modified by  the initial-state interactions of the annihilating antiquark.  
A simple argument for this sign change is given in ref.~\cite{Brodsky:2013oya}

Similarly, the final-state interactions of the produced quark with its comoving spectators in SIDIS produces a final-state $T$-odd polarization correlation -- the ``Collins effect".  This can be measured without beam polarization by measuring the correlation of the polarization of a hadron such as the $\Lambda$ baryon with the quark-jet production plane.  Analogous spin effects occur in QED reactions due to the rescattering via final-state Coulomb interactions. Although the Coulomb phase for a given partial wave is infinite, the interference of Coulomb phases arising from different partial waves leads to observable effects.  These considerations have led to a reappraisal of the range of validity of the standard factorization ansatz.~\cite{Collins:2007nk}

The calculation of the Sivers single-spin asymmetry in deep inelastic lepton scattering in QCD requires two different orbital angular momentum components: $S$-wave with the quark-spin parallel to the proton spin and $P$-wave for the quark with anti-parallel spin; the difference between the final-state ``Coulomb" phases leads to a $\vec S \cdot \vec q \times \vec p$ correlation of the proton's spin with the virtual photon-to-quark production plane.~\cite{Brodsky:2002cx}  Thus, as it is clear from its QED analog,  the final-state gluonic interactions of the scattered quark lead to a  $T$-odd non-zero spin correlation of the plane of the lepton-quark scattering plane with the polarization of the target proton.~\cite{Brodsky:2002cx}

The  $S$- and $P$-wave proton wavefunctions also appear in the calculation of the Pauli form factor quark-by-quark. Thus one can correlate the Sivers asymmetry for each struck quark with the anomalous magnetic moment of the proton carried by that quark,~\cite{Lu:2006kt}  leading to the prediction that the Sivers effect is larger for  positive pions as seen by the 
HERMES experiment at DESY,~\cite{Airapetian:2004tw} the COMPASS experiment~\cite{Bradamante:2011xu,Alekseev:2010rw,Bradamante:2009zz} at CERN, and CLAS at Jefferson Laboratory~\cite{Avakian:2010ae,Gao:2010av}
The physics of the ``lensing dynamics" or Wilson-line physics~\cite{Brodsky:2010vs} underlying the Sivers effect  involves nonperturbative quark-quark interactions at small momentum transfer, not  the hard scale $Q^2$  of the virtuality of the photon.  It would interesting to see if the strength of the 
soft initial- or final- state scattering can be predicted using the effective confining potential of QCD from light-front holographic QCD.

Measurements~\cite{Falciano:1986wk} of the Drell-Yan Process $\pi p \to \mu^+ \mu^- X$ display an angular distribution which contradicts pQCD expectations.     In particular one observes an anomalously large $\cos{ 2\phi} $ azimuthal angular correlation between the lepton decay plane and its production plane which contradicts the Lam-Tung relation, a prediction of perturbative QCD factorization.~\cite{Lam:1980uc}  Such effects again point to the importance of initial and final-state interactions of the hard-scattering constituents,~\cite{Boer:2002ju}  corrections not included in the standard pQCD factorization formalism.  For example, if both the
quark and antiquark in the Drell-Yan subprocess
$q \bar q \to  \mu^+ \mu^-$ interact with the spectators of the
other  hadron, then one predicts a $\cos 2\phi \sin^2 \theta$ planar correlation in unpolarized Drell-Yan
reactions.~\cite{Boer:2002ju}  This ``double Boer-Mulders effect" can account for the anomalously large $\cos 2 \phi$ correlation observed by the NA10 collaboration~\cite{Falciano:1986wk}  and the violation~\cite{Boer:2002ju, Boer:1999mm} of the Lam Tung relation  a standard prediction based on perturbative QCD factorization.~\cite{Lam:1980uc}    Such effects point to the importance of both initial and final-state interactions of the hard-scattering constituents, corrections not included in the standard pQCD factorization formalism. 
One also observes large single-spin asymmetries in reactions such as $ p p_\updownarrow \pi X$, an effect not yet explained.~\cite{Liang:1993rz} Another important signal for factorization breakdown at the LHC  will be the observation of a $\cos 2 \phi$ planar correlation in dijet production.
As  emphasized by Collins and Qiu,~\cite{Collins:2007nk} the traditional factorization formalism of perturbative QCD  fails in detail for many hard inclusive reactions because of initial- and final-state interactions.

The final-state interactions of the struck quark with the spectators~\cite{Brodsky:2002ue}  also lead to diffractive events in deep inelastic scattering (DDIS) at leading twist,  such as $\ell p \to \ell' p' X ,$ where the proton remains intact and isolated in rapidity;    in fact, approximately 10 \% of the deep inelastic lepton-proton scattering events observed at HERA are
diffractive.~\cite{Adloff:1997sc, Breitweg:1998gc} This seems surprising since the underlying hard subprocess $\ell q \to \ell^\prime q^\prime$ is highly disruptive of the target nucleon.
The presence of a rapidity gap
between the target and diffractive system requires that the target
remnant emerges in a color-singlet state; this is made possible in
any gauge by the soft rescattering incorporated in the Wilson line or by augmented light-front wavefunctions. 
Quite different fractions of single $ p p \to {\rm Jet}  \,p^\prime  X$ and double diffractive
$p \bar p \to {\rm Jet} \, p^\prime  \bar p^\prime X$ events are observed at the Tevatron. The underlying mechanism is believed to be soft gluon exchange between the scattered quark and the remnant system in the final state occurring after the hard scattering  occurs.

\section{The Principle of Maximum Conformality}

It is conventional to 
guess the renormalization scale $\mu$  of the QCD coupling $\alpha_s(\mu^2)$  and its range in pQCD predictions.
The resulting predictions then depend on the choice of the renormalization scheme.
This arbitrary procedure of guessing the renormalization scale and range violates the principle of 
``renormalization group invariance":  physical observables cannot depend on the choice of the renormalization scheme or the initial scale.  Varying the renormalization scale can only expose terms in the pQCD series which are proportional to the $\beta$ function; it is thus an unreliable way to estimate the accuracy of pQCD predictions. 

In fact, there is a rigorous method -- the ``Principle of Maximum Conformality" (PMC)~\cite{Brodsky:2013vpa,Brodsky:2011ig} which systematically eliminates the renormalization scale uncertainty at each order of perturbation theory.   The PMC provides scheme-independent predictions at  each finite order in perturbative QCD by systematically identifying and shifting the nonconformal $\beta\ne 0$ terms into the QCD running coupling.   The renormalization scale at each order of perturbation theory is thus fixed by systematically identifying and resuming the nonconformal $\beta$ terms into the QCD running coupling.  

The PMC is a rigorous extension of the BLM method~\cite{Brodsky:1982gc}.  It reduces in the Abelian limit~\cite{Brodsky:1997jk} $N_C\to 0$ at fixed $C_F\alpha_s =\alpha$ to the standard  Gell Mann-Low method for setting the renormalization scale in precision QED predictions;  in the case of QED, the renormalization scale is simply the photon virtuality.  The PMC also sets the effective number of active flavors $n_F$ appearing in loops correctly at each order of perturbation theory, and it can be applied to multi-scale problems. 

The identification of the $\beta$ terms at any order in pQCD can be performed unambiguously using the $R_\delta$ method~\cite{Mojaza:2012mf}, which systematically  exposes all of the $\beta_i$ terms in any dimensional regularization scheme. One first generalizes the $\overline {MS}$ scheme  by subtracting a constant $\delta$ in the dimensional regularization of the UV divergent amplitudes which renormalize the QCD running coupling, in addition to the usual  $log 4\pi-\gamma_E $ subtraction. The resulting terms in $\delta$ reveal the pattern of nonconformal $\beta_i\ne 0$ terms.  The renormalization scales are then fixed order-by-order by shifting the arguments of $\alpha^n_s,$ so that no terms in $\delta$ or $\beta_i$ appear.
The coefficients of the resulting series matches that of the corresponding scheme-independent $\beta=0$ conformal theory. The PMC predictions are thus independent of the choice of renormalization scheme, as required by renormalization group invariance. All other principles of the renormalization group, 
such as transitivity and reciprocity, are satisfied~\cite{Brodsky:1994eh}.
The  divergent   ``renormalon" $\alpha^n_s \beta_0^n n !  $ terms in the pQCD series based on a guessed scale are also eliminated.  
These are also the principles underlying the scheme-independent commensurate scale relations~\cite{Brodsky:1994eh} between observables such as the Generalized Crewther Relation~\cite{Brodsky:1995tb}

\begin{figure}
 \begin{center}
\includegraphics[height=6cm,width=8cm]{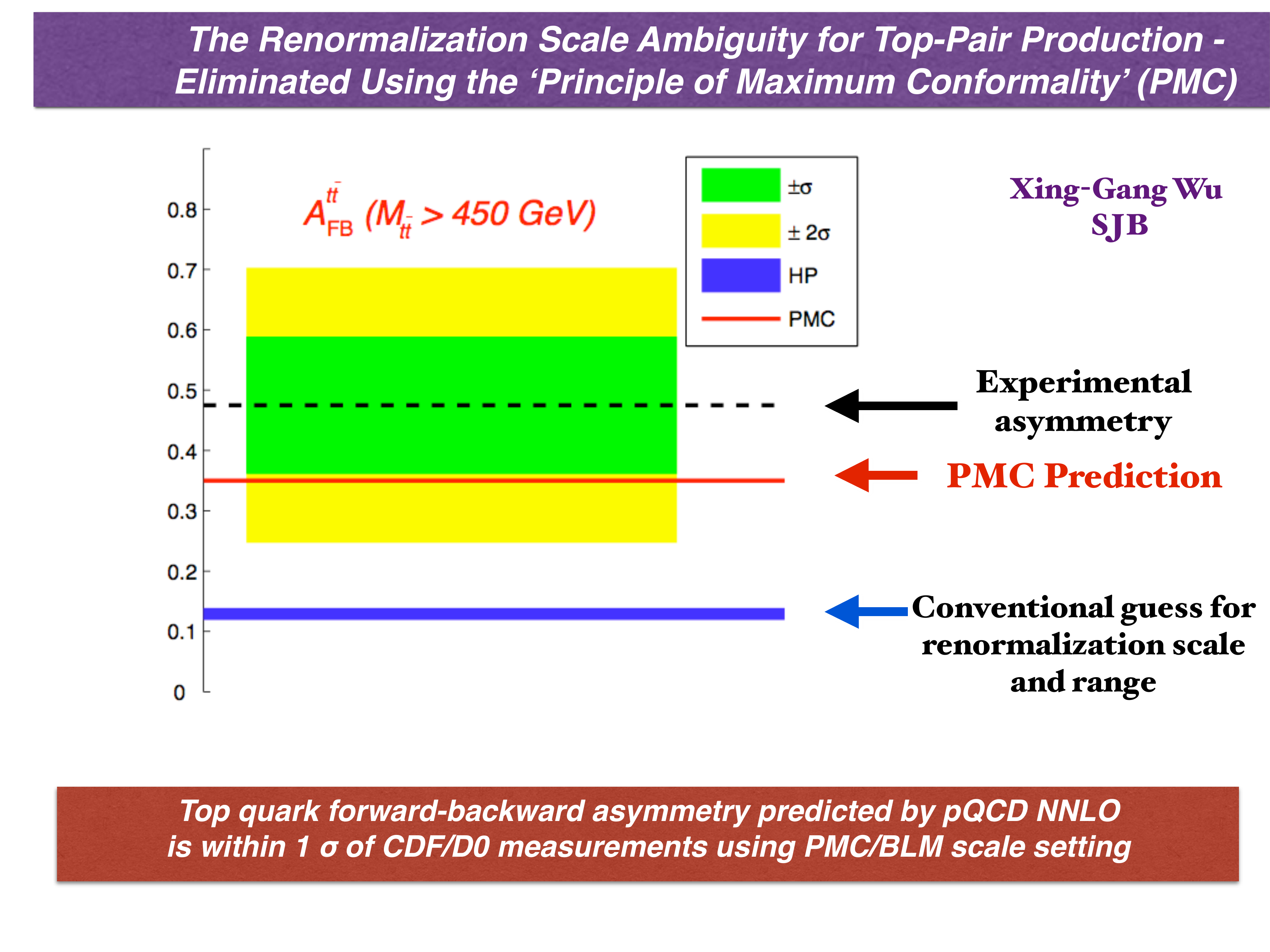}
\end{center}
\caption{Application of the PMC to $p \bar p \to t \bar t X$ forward-backward asymmetry..} 
\label{Top}  
\end{figure}

In the case of the forward-backward asymmetry in  $\bar p p \to t \bar t X$ at the Tevatron, the application of the PMC~\cite{Brodsky:2012ik}
eliminates the anomaly reported by CDF and D0 -- the discrepancy  between measurements and pQCD predictions was based on the choice of an erroneous choice of renormalization scale and range.  See fig. \ref{Top}
As in the QED analog,   $e^+ e^- \to \mu^+ \mu^-$,  the higher  Born  amplitudes which produce  the $\mu^+ \mu^-$ forward backward asymmetry have a smaller renormalization scale than the lowest-order amplitude.  Thus it is essential to assign a different renormalization scale at each order of perturbation theory.  The effective number of flavors $n_f$ is also different at each order.  

As we have recently shown~\cite{Brodsky:2012sz},  the scale uncertainty for top-pair production at the LHC can be eliminated at the NNLO level.  
When one includes the known higher-order contributions, the central values predicted by the PMC become closer to the central value of the LHC measurements.  The application of the PMC scale setting  to Higgs branching ratios is given in Ref.~\cite{Wang:2013bla}. 

The PMC thus provides a systematic and unambiguous way to set the renormalization scale of any process at each order of PQCD.  An unnecessary  error from theory is eliminated.  The PMC thus allows the LHC to test QCD much more precisely, and the sensitivity of LHC measurements to physics beyond the Standard Model is greatly increased. The PMC is clearly an important advance for LHC physics since it provides an important opportunity to strengthen tests of fundamental theory.

\section*{Acknowledgements}
I am grateful to my collaborators and to  Stephan Narison for his invitation to Montpellier and for organizing QCD2014.   
This research was supported by the Department of Energy  contract DE--AC02--76SF00515.  
SLAC-PUB-16079

\nin



\end{document}